\documentclass[12pt,a4paper]{article}
\usepackage{amssymb,amsmath,amsthm,latexsym}
\usepackage{mathrsfs}
\usepackage[english]{babel}
\usepackage[latin1]{inputenc}
\usepackage{amsfonts}
\usepackage{graphics}

\usepackage[pagebackref]{hyperref}

\def \F {\mathbb F}

\def\balpha{\boldsymbol{\alpha}}
\def\bbeta{\boldsymbol{\beta}}

\def\lm{LM}

\def \Pp {\mathcal{P}}

\def\X{\mathcal{X}}

\def\aff{\textrm{Aff\/}}

\newtheorem{theorem}{Theorem}[section]
\newtheorem{corollary}[theorem]{Corollary}
\newtheorem{proposition}[theorem]{Proposition}
\newtheorem{definition}[theorem]{Definition}

\newtheorem{lemma}[theorem]{Lemma}

\begin{document}

\begin{center}
{\Large\textbf{A family of codes with locality containing optimal codes}}
\end{center}
\vspace{3ex}

\noindent\begin{center} 
\textsc{Bruno Andrade$^{1}$, C\'{\i}cero  Carvalho$^{1}$, Victor G.L. 
Neumann$^{1}$ and 
Ant\^{o}nio C.P.\ Veiga$^{2}$}\\ 
\vspace{1ex}
\small{$^{1}$Faculdade de Matem\'{a}tica, $^{2}$Faculdade de Engenharia 
El\'{e}trica \\ Universidade Federal de 
Uberl\^{a}ndia, 
Av. 
J. N. 
\'{A}vila  2121, 38.408-902 Uberl\^{a}ndia -- MG, Brazil }
\end{center}

\vspace{8ex}

\vspace{4ex}
\begin{small}
\begin{center}
\textbf{Abstract}
\end{center}
Locally recoverable codes were introduced by Gopalan et al.\ in 2012, and in 
the same year Prakash et al.\ introduced the concept of codes with locality, 
which are a type of locally recoverable codes. In this work we introduce a new 
family of codes with locality, which are subcodes of a certain family of 
evaluation codes. We determine the dimension of these codes, and also bounds 
for the minimum distance. We present the true values of the minimum distance in 
special cases, and also show that elements of this family are ``optimal 
codes'', as defined by Prakash et al.
\end{small}

%\title{Locally Recoverable Affine Cartesian Codes}
%
%\author{Bruno Andrade, C\'{\i}cero  Carvalho and Victor G.L. Neumann}
%\date{\today}
%%\date{04  de Setembro de 2018}
%\subjclass[2010]{11T71,94B27,14G50}
%\keywords{Locally Recoverable Codes,  Affine Cartesian Codes}
%
%\begin{abstract}
%To see
%\end{abstract}
%
%\maketitle

\section{Introduction}
The class of locally recoverable codes was introduced in 2012 by Gopalan et 
al.\ 
(see \cite{lrc1}). The idea was to ensure reliable communication when using 
distributed storage systems. Thus the authors define a code as having locality 
$r$ if an entry  at position $i$ of a codeword of length $n$ may be recovered 
from a set (which may vary with $i$) of at most $r$ other entries, for all $i = 
1, \ldots, n$. This would ensure the recovering of a codeword even in the 
presence of an erasure, due for example to a failure of some node in the 
network. In that same year Prakash et al.\ (see \cite{prakash}) introduced the 
concept of codes with 
locality $(r, \delta)$, also called $(r, \delta)$-locally recoverable codes,  
which are codes of length $n$ such that for every 
position $i \in \{1, \ldots, n\}$ there is a subset $S_i \subset \{1, \ldots, 
n\}$ containing $i$ and of size at most $r + \delta - 1$ such that the $i$-th 
entry of a codeword may be recovered from any subset of $r$ entries with 
positions in $S_i\setminus \{i\}$, so that we may recover any entry even with 
$\delta - 2$ other erasures in the code.

In this paper we define a family consisting of subcodes of the so-called affine 
cartesian 
codes (see Definition \ref{def1}) which are  $(r, \delta)$-locally recoverable 
codes. We call this family quasi affine cartesian codes. We determine their 
dimension (see Corollary  \ref{dim-dtilda} and 
Theorem \ref{dimensao}) together with lower 
and upper bounds for the minimum distance (see  Theorem \ref{min_dist_1}). 
We list some cases where the codes are optimal (see Corollary \ref{optimal}) and 
we also determine the exact value of the minimum distance in some special cases 
of the code (see Theorem \ref{min_dist_1}, Theorem \ref{secondweight} and 
Corollary \ref{last-result}). 

In the next section we introduce the family of quasi affine cartesian codes, 
and prove that these codes are locally recoverable.  In Section 3 we present 
several results on the dimension of these codes, after recalling some facts 
from Gr\"obner basis theory which we will need. In the following section we 
present lower and upper bounds for the minimum distance of quasi affine codes, 
and determine the exact values in some cases. We also prove that some of the 
codes we introduced are optimal codes. In Section 5 we treat a special 
case of quasi affine cartesian codes, for which we determine more values for 
the minimum distance. The paper ends with several numerical examples.

\section{Quasi affine cartesian codes}
Let $\F_q$ be a finite field with $q$ elements. 

\begin{definition}
Let $m, r, \delta$ be positive integers, with $\delta \geq 2$ and $r + \delta - 
1 \leq m$. We say that 
a (linear) code $\mathcal{C} \subset \mathbb{F}_q^m$ is  $(r, 
\delta)$-locally recoverable if for every $i \in \{1, \ldots, m\}$ there exists 
a subset $S_i \subset \{1, \ldots, n\}$, containing $i$ and of cardinality at 
most $r + \delta - 1$, such that the punctured code obtained by removing the 
entries which are not in $S_i$ has minimum distance at least $\delta$.
\end{definition}

The condition on the minimum distance in the above definition shows that one 
cannot have two distinct codewords in the punctured code which coincide in (at 
least) $r$ 
positions, so any $r$ positions in the set $S_i$  determine the remaining 
$\delta - 1$ positions.

Let $K_1,\ldots, K_n$ be 
a collection of non-empty subsets of $\F_q$, and let 
$$
\mathcal{X}:= K_1\times\cdots\times K_n:=\left\{(\alpha_1,\ldots ,\alpha_n) \vert\, \alpha_i\in
K_i
\mbox{ for all } i\right\}\subset \F_q^n.
$$

Let $d_i := | K_i |$ for $i = 1,\ldots, n$, so 
clearly $| \mathcal{X}| = \prod_{i = 1}^n d_i =: m$, and let 
$\mathcal{X} = \{\balpha_1, \ldots, \balpha_m\}$. It is not difficult to check that the
ideal of polynomials in $\F_q[X_1, \ldots, X_n]$ which vanish on $\mathcal{X}$ is
$$
I_\mathcal{X} = 
\left(
\prod_{\alpha_1 \in K_1} (X_1 - \alpha_1),\ldots ,  \prod_{\alpha_n \in K_n} (X_n - \alpha_n)
\right)
$$
(see e.g.
\cite[Lemma 2.3]{lopez-villa} or \cite[Lemma 3.11]{car2}). The evaluation 
morphism 
\[
\begin{array}{ccc}
\Psi \colon \F_q[X_1, \ldots, X_n] & \rightarrow  & \F_q^m \\ 
f & \mapsto & ( f(\balpha_1), \ldots, 
f(\balpha_m) )
\end{array} 
\]
is an
$\F_q$-linear map and $\ker \Psi = I_{\X}$.
Actually, this is a surjective map  because for
each $i \in \{1, \ldots, m\}$ there exists a  polynomial $f_i$ such that
$f_i(\balpha_j)$ is equal to $1$, if $j = i$, or $0$, if $j \neq i$.

Let $d$ be a nonnegative integer.
In what follows we will denote by $\F_q[X_1, \ldots, X_n]_{\leq d}$ the 
$\F_q$-vector space formed by all polynomials of degree up to $d$, together 
with 
the zero polynomial.

\begin{definition}  \label{def1}
Let $d$ be a nonnegative integer. 
The {\em affine cartesian code} (of order $d$) $\mathcal{C}_\mathcal{X}(d)$
defined over the sets $K_1, \ldots, K_n$ is the image, by $\Psi$, of the 
polynomials in $\F_q[X_1, \ldots, X_n]_{\leq d}$.
\end{definition}  

These codes appeared
independently  in \cite{lopez-villa} and \cite{Geil} (in \cite{Geil}  in a 
generalized form). In the special case
where $K_1 = \cdots = K_n = \F_q$ we have the well-known generalized Reed-Muller
code of order $d$. In \cite{lopez-villa} the authors prove that we may ignore,
in the cartesian product, sets with just one element and moreover may always
assume that $2 \leq d_1 \leq \cdots \leq d_n$. The dimension
and the minimum distance of these codes are known (see e.g.\ \cite{lopez-villa} 
or \cite{Geil}).

In what follows we construct $(r,\delta)$-locally recoverable codes 
which are subcodes
of affine cartesian codes.

\begin{definition}  \label{def2}
Let $d$ and $\delta$ be integers with $d \geq 0$ and 
$\delta \geq 2$,
let $s \in \{1,\ldots , n \}$ and
let $\Pp^{(\delta, s)}_d$ be the set of polynomials
$f \in \F_q[X_1,\ldots ,X_n]_{\leq d}$
such that $\deg_{X_{s}} f < d_s-\delta+1$, together with
the zero polynomial.
The {\em $(\delta,s)$-quasi affine cartesian code} (of order $d$) 
$\mathcal{D}^{(\delta,s)}_\mathcal{X}(d)$
defined over the sets $K_1, \ldots, K_n$ is the image, by $\Psi$, of the set
$\Pp^{(\delta,s)}_d$.
\end{definition}

%Observe that $\mathcal{D}^{(\delta,j)}_\mathcal{X}(d) \subset \mathcal{C}_\mathcal{X}(d)$
%and we have equality if and only if $d \leq d_j - \delta$
%because $\Psi (X_j^{d_j-\delta+1}) \notin \Psi (\Pp^{(\delta,j)}_d)$ for any $d\geq 
%d_j-\delta+1$.

\begin{theorem}\label{code_is_locally_r}
Let $K_1,\ldots , K_n$ be  subsets of $\F_q$ such that $|K_i|=d_i \geq 2$ for
all $i=1,\dots,n$, with $n \geq 2$, let $\delta \geq 2$ be an integer
such that $d_s - \delta + 1 \geq 1$ and let 
$d$ be a nonnegative integer.
For any $s \in \{1 
,\ldots , n\}$,
the {\em $(\delta,s)$-quasi affine cartesian code} (of order $d$) 
$\mathcal{D}^{(\delta,s)}_\mathcal{X}(d)$
defined over the sets $K_1, \ldots, K_n$ is locally recoverable with locality
$(r,\delta)$ where $r=d_s-\delta+1$.
\end{theorem}
\begin{proof}
Let $f \in \Pp^{(\delta, s)}_d$, so $( f(\balpha_1), \ldots, f(\balpha_m) )
\in \mathcal{D}^{(\delta,s)}_\mathcal{X}(d)$. Let $\balpha=(\alpha_1,\ldots , 
\alpha_n) \in \mathcal{X}$ and let 
$$
I_{\balpha} = \{ (\alpha_1,\ldots \alpha_{s-1},\beta,
\alpha_{s+1},\ldots  \alpha_n) \mid \beta \in K_s \},
$$
a set which has $d_s = r+\delta - 1$ elements. Assume 
that there  exist
$\bbeta_1,\ldots ,\bbeta_r \in I_{\balpha}$ such that we know the values
$f(\bbeta_k) =: c_k$, for $k \in \{1,\ldots , r\}$, we will prove that then 
we can deduce the value of $f(\bbeta)$ for any $\bbeta \in I_{\balpha}$.

Write
$f=\sum_{i=0}^{r-1} g_i X_s^i$,
where $g_1, \ldots, g_{r - 1}$ are polynomials in the variables $X_1,\ldots , 
X_{s-1},X_{s+1},\ldots ,X_n$, and let $b_i := g_i(\alpha_1,\ldots 
\alpha_{s-1},\alpha_{s+1},\ldots  \alpha_n)$ for $i = 0, \ldots, r - 1$.
Denoting by $\beta_k$ the $s$-th coordinate of $\bbeta_k$, for $k = 1, \ldots, 
r$,
from the assumption we get that
$$
c_k = f(\bbeta_k) = 
\sum_{i=0}^{r-1} b_i \beta_k^i, \quad \text{for } k \in \{ 1,\ldots , r\}.
$$
This system of equations can be rewritten as a matrix equation 
$$
\left(
\begin{matrix}
1 & \beta_1 & \beta_1^2 & \cdots & \beta_1^{r-1} \\
1 & \beta_2 & \beta_2^2 & \cdots & \beta_2^{r-1} \\
1 & \beta_3 & \beta_3^2 & \cdots & \beta_3^{r-1} \\
1 & \colon & \colon & \ddots & \colon \\
1 & \beta_r & \beta_r^2 & \cdots & \beta_r^{r-1} 
\end{matrix}
\right)
\left(
\begin{matrix}
b_0 \\
b_1 \\
b_2\\
\colon \\
b_{r-1}
\end{matrix}
\right)=
\left(
\begin{matrix}
c_1 \\
c_2 \\
c_3 \\
\colon \\
c_{r}
\end{matrix}
\right),
$$
which has a unique solution $(b_0,b_1,\ldots ,b_{r-1})$, since
the square $r \times r$ matrix is a Vandermonde matrix.
This allow us to determine $f(\bbeta)$ for any $\bbeta \in I_{\balpha}$.
\end{proof}

\section{On the dimension of quasi affine cartesian codes}\label{footprint}

In this section we determine the dimension of $(\delta,s)$-quasi affine 
cartesian codes, and we will need some facts about Gr\"{o}bner basis which we 
recall below.

Let $\prec$ be a monomial order in (the set of monomials of) 
$\F_q[X_1,\ldots,X_n]$, i.e.\ $\prec$ is a total order, if $M_1 
\prec M_2$ then $M M_1  \prec M M_2$ for all monomials $M, M_1, M_2$, and 1 is 
the least monomial. The greatest monomial appearing in a polynomial $f$ is 
called the leading monomial of $f$ and is denoted by $\lm(f)$.

\begin{definition} 
Let $J \subset  \F_q[X_1,\ldots,X_n]$ be an ideal. A {\em Gr\"obner basis} 
(with respect to a monomial order $\prec$) for $J$ is a basis $G$ for $J$ such 
that the leading 
monomial of any polynomial in $J$ is a multiple of the leading monomial of some 
polynomial in $G$. The 
{\em footprint}  of $J$ 
(with respect to a monomial order $\prec$) is the set of monomials  of 
$\F_q[X_1,\ldots,X_n]$ 
which are not leading monomials of any polynomials in $J$, and is denoted by 
$\Delta(J)$.
\end{definition}

B.\ Buchberger proved that, given a monomial order, any (nonzero) ideal $J 
\subset  \F_q[X_1,\ldots,X_n]$  admits a Gr\"obner basis (see \cite{bruno} or 
\cite[Sec.\ 1.7]{adams}). He also proved that a basis for 
$\F_q[X_1,\ldots,X_n]/J$ as an $\F_q$-vector space is given by the classes of 
the monomials in $\Delta(J)$ (see e.g.\ \cite[Prop. 2.1.6]{adams}).  

\begin{definition}
Let $\prec$ be a monomial order in $\F_q[X_1,\ldots,X_n]$ and 
let $J \subset  \F_q[X_1,\ldots,X_n]$ be an ideal.
Let $\{g_1, \ldots, g_r\}$ be a (not necessarily Gr\"obner) basis for $J$, 
we define $\Delta( \lm(g_1), \ldots, \lm(g_r) )$ as the set of monomials of 
$\F_q[X_1,\ldots,X_n]$
which are not multiples of any of the leading monomials of $g_1, \ldots, g_r$.
\end{definition}

Clearly we have $\Delta(J) \subset \Delta( \lm(g_1), \ldots, \lm(g_r) )$ and, 
moreover, $\Delta(J) = \Delta( \lm(g_1), \ldots, \lm(g_r) )$ if and only if 
$\{g_1,\ldots, g_r\}$ is a Gr\"obner basis for $J$.
 
%
%
%We will denote by $\Delta(J)$ the footprint of $J$. 
%A well-known property of the footprint is that the classes of the elements of 
%$\Delta(J)$ are a basis for $\F_q[X_1,\ldots,X_n]/J$ as a $K$-vector space 
%(see 
%e.g.\ \cite[Prop.\ 6.52]{becker}).
%For a nonzero polynomial $g \in \F_q[X_1,\ldots,X_n]$ we denote by $\lm(g)$ 
%the leading monomial of $g$. Let $J = \langle g_1,\ldots, g_r \rangle$ and let 
%$\Delta( \lm(g_1), \ldots, \lm(g_r) )$ be the set of monomials of 
%$\F_q[X_1,\ldots,X_n]$
%which are not a multiple of the leading monomial of $g_i$ for all $i \in \{1, 
%\ldots, r\}$, then $\Delta(J) \subset \Delta( \lm(g_1), \ldots, \lm(g_r) )$.  
%From the definition of Gr\"obner basis (with respect to $\prec$) we get that a 
%monomial is in
%$\Delta(J)$ if and only if it is not a multiple of any of the leading 
%monomials of the polynomials in a Gr\"obner basis for $J$ (see e.g. 
%\cite[Prop. 
%2.12]{car2}),
%so that $\Delta(J) = \Delta( \lm(g_1), \ldots, \lm(g_r) )$ if and only if 
%$\{g_1,\ldots, g_r\}$ is a Gr\"obner basis for $J$.

In what follows we will use
the graded-lexicographic order in $\F_q[X_1,\ldots,X_n]$, with $X_n \prec 
\cdots \prec X_1$.

For $i = 1, \ldots, n$ let 
$f_i = \prod_{\alpha \in K_i} (X_i - \alpha)$, so that
$\deg f_i = d_i$ and $I_\mathcal{X}= \langle f_1 , \ldots , f_n \rangle$. 
Since any two of the leading monomials of $f_1, \ldots, f_n$ are coprime we get that $\{f_1, \ldots, f_n\}$ is a Gr\"obner basis for $I_\mathcal{X}$ (see \cite[Prop. 4, page 104]{iva}) so 
$$
\Delta(I_\mathcal{X}) = \Delta( X_1^{d_1}, \ldots, X_n^{d_n}) =
\left\{
X_1^{a_1}\cdots X_n^{a_n} \mid 0 \leq a_i < d_i \, , \, \forall i=1,\ldots ,n
\right\}.
$$ 
Let $\Delta(I_{\mathcal{X}})_{\leq d} = 
\{ M \in \Delta(I_{\mathcal{X}}) \mid \deg (M) \leq d\}$,
it is known (see e.g.\ \cite[Prop.\ 3.12]{car2}) that 
$\dim(\mathcal{C}_{\mathcal{X}}(d)) = |\Delta(I_{\mathcal{X}})_{\leq d}|$. This 
implies that if $d \geq  \sum_{i = 1}^n (d_i - 1)$ then  
$\dim(\mathcal{C}_{\mathcal{X}}(d)) = 
|\Delta(I_{\mathcal{X}})_{\leq d}| = |\Delta(I)| =  \prod_{i = 1}^n d_i$, while 
if $0 \leq d < \prod_{i = 1}^n d_i$ then 
\begin{equation*}
\begin{split}
  & \dim(\mathcal{C}_{\mathcal{X}}(d)) =  \binom{n + d }{d}  -   \sum_{i = 1}^n 
  \binom{n + d - d_i}{d - 
  d_i} + \cdots +   \\ & (-1)^j \sum_{1 \leq i_1 < \cdots < i_j \leq n} 
  \binom{n + d - d_{i_1} - \cdots -   d_{i_j}}{d - d_{i_1} - \cdots -   
  d_{i_j}} + \cdots +   (-1)^n \binom{n + d - d_{1} - \cdots -   d_{n}}{d - 
  d_{1} - \cdots -   d_{n}} 
\end{split}
\end{equation*}
where we set $\binom{a}{b} = 0$ if $b < 0$.
 
To determine the dimension of $\mathcal{D}^{(\delta,s)}_\mathcal{X}(d)$ we  
make a reasoning similar to the one used to prove the above formulas.

\begin{proposition}
Let
$\Delta(I_{\mathcal{X}})_{\leq d}^{(\delta, s)} = \{ M \in 
\Delta(I_{\mathcal{X}})_{\leq d} \mid \deg_{X_{s}} M < d_s - \delta + 1\}$,
then $\dim(\mathcal{D}^{(\delta,s)}_\mathcal{X}(d)) = | 
\Delta(I_{\mathcal{X}})_{\leq d}^{(\delta, s)} |$.
\end{proposition}
\begin{proof}
Given $f \in \Pp^{(\delta, s)}_d$ let $g\in \F_q[X_1,\ldots , X_n]$
be its remainder in the division by
$\{ f_1,\ldots ,f_n\}$, then $\Psi(f)=\Psi(g)$. From the division algorithm 
we know that any monomial which appear in $g$ is not a multiple of $\lm(f_i) = 
X_i^{d_i}$ for all $i = 1, \ldots, n$, and also that  
$\deg g \leq \deg f$  and  $\deg_{X_{s}} g < d_s-\delta+1$. Thus $g \in 
\Pp^{(\delta, s)}_d$ and moreover, $g$ is a 
linear combination of monomials in $\Delta(I_{\mathcal{X}})_{\leq d}^{(\delta, 
s)}$. This shows that $\dim(\mathcal{D}^{(\delta,s)}_\mathcal{X}(d)) \leq | 
\Delta(I_{\mathcal{X}})_{\leq d}^{(\delta, s)} |$. Let 
\[
\overline{\Psi} \colon 
\F_q[X_1, \ldots, X_n]/I_{\mathcal{X}} \rightarrow \F_q^m
\]
be defined as $\overline{\Psi}(f + 
I_{\mathcal{X}}) = \Psi(f)$, we know that $\overline{\Psi}$ is an isomorphism 
and clearly $\mathcal{D}^{(\delta,s)}_\mathcal{X}(d) = \{ \overline{\Psi}(h + 
I_{\mathcal{X}}) \mid h \in \langle \Delta(I_{\mathcal{X}})_{\leq d}^{(\delta, 
s)}  \rangle \}$, where $\langle \Delta(I_{\mathcal{X}})_{\leq d}^{(\delta,
s)}  \rangle$ is the $\F_q$-vector space generated by the monomials in 
$\Delta(I_{\mathcal{X}})_{\leq d}^{(\delta,
s)}$. Since $\Delta(I_{\mathcal{X}})_{\leq d}^{(\delta,s)} \subset  
\Delta(I_{\mathcal{X}})$ we know from Buchberger's result that the classes 
in $\F_q[X_1, \ldots, X_n]/I_{\mathcal{X}}$ of the monomials in 
$\Delta(I_{\mathcal{X}})_{\leq d}^{(\delta,s)}$  are linearly independent over 
$\F_q$, 
thus we get  $\dim(\mathcal{D}^{(\delta,s)}_\mathcal{X}(d)) = | 
\Delta(I_{\mathcal{X}})_{\leq d}^{(\delta, s)} |$.
\end{proof}

Let $\tilde{d} := \displaystyle \sum_{\substack{i=1 \\ i \neq s}}^n (d_i -1) 
+d_s - \delta$.

\begin{corollary}\label{dim-dtilda}
If $d \geq \tilde{d}$ then $\mathcal{D}^{(\delta,s)}_\mathcal{X}(d)= 
\mathcal{D}^{(\delta,s)}_\mathcal{X}(\tilde{d})$, and 
\[
\dim(\mathcal{D}^{(\delta,s)}_\mathcal{X}(\tilde{d})) = (d_s - \delta +1) 
\prod_{\substack{i=1 \\ i \neq s}}^n d_i.
\]
Also $\dim(\mathcal{D}^{(\delta,s)}_\mathcal{X}(\tilde{d} - 1)) = 
\dim(\mathcal{D}^{(\delta,s)}_\mathcal{X}(\tilde{d})) - 1$.
\end{corollary}
\begin{proof} 
From the above proof we get that if $M \in \Delta(I_{\mathcal{X}})_{\leq 
d}^{(\delta, s)}$ then $\deg_{X_{s}}(M) < d_s - \delta + 1$ and    
$\deg_{X_{i}}(M) < d_i$ for all $i \in \{1, \ldots, n\} \setminus \{s\}$.
Thus if $d \geq \tilde{d}$ we have $\Delta(I_{\mathcal{X}})_{\leq 
d}^{(\delta, s)} = \Delta(I_{\mathcal{X}})_{\leq 
\tilde{d}}^{(\delta, s)}$ which implies 
$\mathcal{D}^{(\delta,s)}_\mathcal{X}(d)= 
\mathcal{D}^{(\delta,s)}_\mathcal{X}(\tilde{d})$. We also have that  
\begin{equation*}
\begin{split}
&\dim(\mathcal{D}^{(\delta,s)}_\mathcal{X}(\tilde{d})) \\  &= |\left \{
X_1^{a_1}\cdots X_n^{a_n} \mid 0 \leq a_i < d_i \, , \,  i=1,\ldots ,n, 
i \neq s, \textrm{ and } 0 \leq a_s < d_s - \delta + 1
\right\}| \\ &= (d_s - \delta +1) \prod_{\substack{i=1 \\ i \neq s}}^n d_i
\end{split}
\end{equation*}
Observe that in $\Delta(I_{\mathcal{X}})_{\leq 
\tilde{d}}^{(\delta, s)}$ there is only one monomial of degree $\tilde{d}$, so 
that 
$\dim(\mathcal{D}^{(\delta,s)}_\mathcal{X}(\tilde{d} - 1)) = 
\dim(\mathcal{D}^{(\delta,s)}_\mathcal{X}(\tilde{d})) - 1$.
\end{proof}

%\section{Dimension}
Now, for $1 \leq d < \tilde{d}$ (when $d = 0$ we have 
$\mathcal{D}^{(\delta,s)}(0) \simeq \F_q$), we present a formula for  the 
dimension of 
$\mathcal{D}^{(\delta,s)}(d)$ in terms of the dimension of certain affine 
cartesian codes, 
and for that we introduce some notation.

\begin{definition}
For $s \in \{ 1,\ldots , n\}$ we denote by $\X_s$ the product
\[
\X_s = K_1 \times \cdots \times K_{s-1} \times K_{s+1} \times \cdots \times 
K_n.
\]
\end{definition}

Observe that we may define the affine cartesian code 
$\mathcal{C}_{\X_s}(d)$ 
as in Definition \ref{def1}, except that 
now $\mathcal{C}_{\X_s}(d)$ is defined over the sets $K_1,\ldots 
,K_{s-1},$ $K_{s+1},\ldots ,K_n$.

\begin{theorem}\label{dimensao}
Let $s \in \{ 1,\ldots , n\}$ and let $d$ be an integer such that $1 \leq d < 
\tilde{d}$. If $1 \leq d < r=d_s - \delta+1$ then 
$\dim_{\F_q} \mathcal{D}^{(\delta,s)}_\mathcal{X}(d) = \dim_{\F_q} 
\mathcal{C}_\mathcal{X}(d)$, and if 
$r \leq d \leq \tilde{d}$ then 
\begin{equation}\label{dimension}
\dim_{\F_q} \mathcal{D}^{(\delta,s)}_\mathcal{X}(d) = 
\dim_{\F_q} \mathcal{C}_\mathcal{X}(d)
-
\sum_{i=0}^{\delta-2} \dim_{\F_q} \mathcal{C}_{\X_s}(d- r-i),
\end{equation}
where $\dim_{\F_q} \mathcal{C}_{\X_s}(d- r-i) = 0$ if $d- r-i < 0$.
\end{theorem}
\begin{proof}
If $1 \leq d < r = d_s - \delta+1$ then from Definitions \ref{def1} and 
\ref{def2} we get that $\dim_{\F_q} \mathcal{D}^{(\delta,s)}_\mathcal{X}(d) = 
\dim_{\F_q} 
\mathcal{C}_\mathcal{X}(d)$, so we assume now that 
$r \leq d \leq \tilde{d}$. 
Define the following sets:
\begin{eqnarray}
%\Omega & = &
%\{(a_1,\ldots , a_n) \in \N^n \mid 0 \leq a_i < d_i\, , \, \text{for}\, 1 \leq 
%i \leq n
%\};
%\nonumber \\
\Omega_d & = &
\{(a_1,\ldots , a_n) \in \mathbb{N}^n  \mid  0 \leq a_i < d_i\, , \, 
\text{for}\, 1 
\leq 
i \leq n, a_1 + \cdots + a_n \leq d
\};
\nonumber \\
\Omega_d^{(\delta,s)} & = &
\{(a_1,\ldots , a_n) \in \Omega_d  \mid a_s \leq d_s - \delta
\}.
\nonumber 
%\Omega_{s,d-r-i}^{(0)} & = &
%\{(a_1,\ldots , a_n) \in \Omega_{d-r-i} \mid a_s = 0
%\}.
%\nonumber
\end{eqnarray}
From previous considerations we get that 
$\dim_{\F_q} \mathcal{C}_\mathcal{X}(d) = |\Omega_d|$ and 
$\dim_{\F_q} \mathcal{D}^{(\delta,s)}_\mathcal{X}(d)
= |\Omega_d^{(\delta,s)} |$.
For any $(a_1,\ldots , a_n) \in \Omega_d$ we have that either
$a_s \leq d_s - \delta$ or $a_s = d_s - \delta +1 + i$ for
some $i$ in the range $0 \leq i \leq \delta - 2$ (because $a_s \leq d_s - 1$).
If $a_s = d_s - \delta +1 + i = r + i$ then we have 
$$
a_1+\cdots + a_{s-1}+a_{s+1}+\cdots + a_n \leq d - r - i,
$$
and for  $0 \leq i \leq \delta - 2$ we define
\[
\Omega_{s,d-r-i}^{(0)}  = 
\{(a_1,\ldots , a_n) \in \Omega_{d-r-i} \mid a_s = 0
\},
\]
so that $\Omega_{s,d-r-i}^{(0)} = \emptyset$ if 
$i$ is such that $d-r-i < 0$. Thus we have 
$$
|\Omega_d| = |\Omega_d^{(\delta,s)} | +
\sum_{i=0}^{\delta-2} |\Omega_{s,d-r-i}^{(0)}|
$$
and since  $\dim_{\F_q} \mathcal{C}_{\X_s}(d- r-i) = 
|\Omega_{s,d-r-i}^{(0)}|$ for all $i \in \{0, \ldots, \delta - 2\}$ 
the above equation implies
equation \eqref{dimension} in the statement. 
\end{proof}

\section{Minimum distance and optimal codes}

In this section we relate the minimum distance of quasi affine cartesian codes 
to the minimum distance of affine cartesian codes. In what follows we denote by 
$W^{(1)}(C)$ the minimum distance of a code $C$.

Let $d$ be an integer in the range $1 \leq d < \sum_{i = 1}^n (d_i - 1)$, and 
let 
$k$ and $\ell$ be uniquely defined by writing $d = \sum_{i =1}^k (d_i -1) + 
\ell $,  with $0 < \ell \leq d_{k + 1} - 1$ (if $d < d_1 - 1$ then take $k = 0$ 
and $\ell = d$, if $k + 1 = n$ then we understand that $\prod_{i = k + 2}^n d_i 
= 1$). We recall that 
\begin{equation}\label{md-cartesian}
W^{(1)} ( \mathcal{C}_\mathcal{X}(d) )= (d_{k+1}-\ell ) 
\prod_{i=k+2}^n d_i
\end{equation}
(see e.g.\ \cite[Theorem 3.8]{lopez-villa}).

\begin{theorem}	\label{min_dist_1}
	Let $d =\displaystyle \sum_{i=1}^k (d_i-1) + \ell$ where  $ 0 \leq k <n$ and $0 < \ell \leq d_{k+1}-1$.
We have
\begin{multline}\label{minimaldistance}
W^{(1)} ( \mathcal{C}_\mathcal{X}(d) ) \leq 
W^{(1)} ( \mathcal{D}^{(\delta,s)}_\mathcal{X}(d) ) \\
\leq m - 
\dim_{\F_q} \mathcal{D}^{(\delta,s)}_\mathcal{X}(d)
-
\left( \left\lceil   \dfrac{\dim_{\F_q} \mathcal{D}^{(\delta,s)}_\mathcal{X}(d)}{r} \right\rceil - 1 \right)
(\delta -1)
+ 1.
\end{multline}  
where $m=\prod_{i = 1}^n d_i $, $r=d_s-\delta+1$ and for $x \in \mathbb{R}$,
$\lceil x \rceil$ is the smallest integer such that 
$x \leq \lceil x \rceil$.
If
\begin{enumerate}
	\item[(i)] $k+2\leq n$ and $d_{k+2} \leq d_s$, or
	\item[(ii)] $d_s \leq d_{k+1}$ and $0\leq d_s - (d_{k+1} - \ell) < r$
\end{enumerate}
then we get
$W^{(1)} ( \mathcal{D}^{(\delta,s)}_\mathcal{X}(d) )
= W^{(1)} ( \mathcal{C}_\mathcal{X}(d) )$.
\end{theorem}
\begin{proof}
Since $\mathcal{D}^{(\delta,s)}_\mathcal{X}(d)\subset\mathcal{C}_\mathcal{X}(d)$,
we have $W^{(1)} ( \mathcal{C}_\mathcal{X}(d) ) \leq 
W^{(1)} ( \mathcal{D}^{(\delta,s)}_\mathcal{X}(d))$.
From Theorem \ref{code_is_locally_r}
we know that $\mathcal{D}^{(\delta,s)}_\mathcal{X}(d)$
is locally recoverable with locality $(r,\delta)$, so
we may
apply \cite[Theorem 2]{prakash} and we get the second inequality of
\eqref{minimaldistance}.

Assume that $k+2 \leq n$
and $d_{k+2} \leq d_s$. We consider two cases,
$s \geq k+2$ and $s < k+2$, 
let's suppose first that $s \geq k+2$.
Consider an element $\balpha=(\alpha_1,\ldots , \alpha_n) \in \mathcal{X}$,
and consider distinct elements $\beta_1,\ldots ,\beta_{\ell} \in K_{k+1}$.
Define the polynomial
\begin{equation}\label{min_pol}
f=\prod_{i=1}^k \prod_{\substack{\alpha \in K_i \\ \alpha \neq \alpha_i}}
(X_i - \alpha) \cdot \prod_{i=1}^{\ell} (X_{k+1}-\beta_i).
\end{equation}
Observe that $f \in \Pp^{(\delta, s)}_d$
so that $\Psi(f) \in \mathcal{D}^{(\delta,s)}_\mathcal{X}(d)$.
Denoting by $w(v)$ the weight of a codeword $v$ we have
$w(\Psi(f)) = W^{(1)} ( \mathcal{C}_\mathcal{X}(d) )$, and we're done.
Assume now that $s<k+2$, from $d_{k+2} \leq d_s$ we must have 
$K_s=K_{k+1}=K_{k+2}$. Clearly $s \in \{1, \ldots, k+1\}$ so 
replacing $K_s$ by $K_{k+2}$ 
in \eqref{min_pol}
we still have $f \in \Pp^{(\delta, s)}_d$ and 
$w(\Psi(f)) = W^{(1)} ( \mathcal{C}_\mathcal{X}(d) )$. 

Finally suppose that $(ii)$ is satisfied, i.e.\
$s\leq k+1$  and $0 \leq d_s - (d_{k+1} - \ell) < r$, and to avoid
overlapping with the previous case we also assume that either $d_s < d_{k+2}$ 
or $n = k+1$. Now we take
$$
f=\prod_{\substack{i=1\\ i\neq s}}^{k+1}
\prod_{\substack{\alpha \in K_i \\ \alpha \neq \alpha_i}}
(X_i - \alpha) \cdot \prod_{i=1}^{d_s - (d_{k+1} - \ell)} (X_{s}-\beta_i),
$$
where  $\beta_1,\ldots ,\beta_{d_s - (d_{k+1} - \ell)}$ are distinct elements 
of $K_s$, and 
again we have $f \in \Pp^{(\delta, s)}_d$,
$\Psi(f) \in \mathcal{D}^{(\delta,s)}_\mathcal{X}(d)$
and
$w(\Psi(f)) = W^{(1)} ( \mathcal{C}_\mathcal{X}(d) )$.
%So 
%These cases can be reduced to $(i)$ and $(ii)$, where we have
%$W^{(1)}(\mathcal{D}^{(\delta,s)}_\mathcal{X}(d))
%=
%W^{(1)} ( \mathcal{C}_\mathcal{X}(d) )$.
%
%
\end{proof}

Following \cite{prakash}, we say that the code 
$\mathcal{D}^{(\delta,s)}_\mathcal{X}(d)$ is
{\em optimal} if its minimum distance attains the upper bound presented in 
the above theorem.

\begin{corollary} \label{optimal}
The codes $\mathcal{D}^{(\delta,s)}_\mathcal{X}(\tilde{d})$ and 
$\mathcal{D}^{(\delta,s)}_\mathcal{X}(\tilde{d} - 1)$ are optimal, and have 
minimum distance equal to, respectively, $\delta$ and $\delta + 1$.
\end{corollary}
\begin{proof}
We have $\tilde{d} = \displaystyle \sum_{\substack{i=1 \\ i \neq s}}^n (d_i 
-1) +d_s - \delta = \sum_{i=1}^{n-1} (d_i -1) + d_n - \delta$ so from 
\eqref{md-cartesian} we get $W^{(1)} ( \mathcal{C}_\mathcal{X}(\tilde{d}) )
 = d_n - (d_n - \delta)=\delta$. On the other hand, from Corollary 
 \ref{dim-dtilda} and the fact that $r = d_s - \delta + 1$ we get that the upper 
 bound for 
$W^{(1)} ( \mathcal{D}^{(\delta,s)}_\mathcal{X}(\tilde{d}))$ in the above 
theorem is
\begin{multline}
\nonumber
m - 
	\dim_{\F_q} \mathcal{D}^{(\delta,s)}_\mathcal{X}(d)
	-
	\left( \left\lceil   \dfrac{\dim_{\F_q} 
	\mathcal{D}^{(\delta,s)}_\mathcal{X}(d)}{r} \right\rceil - 1 \right)
	(\delta -1)
	+ 1 = \\
\prod_{i=1}^n d_i -
(d_s - \delta +1) \prod_{\substack{i=1 \\ i \neq s}}^n d_i
-
\left(
\prod_{\substack{i=1 \\ i \neq s}}^n d_i - 1
\right)
(\delta -1)
+ 1 = 
\delta 
\end{multline}
so $W^{(1)} ( \mathcal{D}^{(\delta,s)}_\mathcal{X}(\tilde{d})) = \delta$. In 
the same way one proves that $W^{(1)} ( 
\mathcal{D}^{(\delta,s)}_\mathcal{X}(\tilde{d} - 1)) = \delta + 1$.
\end{proof}

One may check that if $d_s=d_n$ and $d \leq \tilde{d}$ then either condition 
$(i)$ 
or condition $(ii)$
of the above Proposition is satisfied, so  we get
$W^{(1)} ( \mathcal{D}^{(\delta,s)}_\mathcal{X}(d) )
= W^{(1)} ( \mathcal{C}_\mathcal{X}(d) )$.
In the following section, among other results, we present some values for 
$W^{(1)} ( \mathcal{D}^{(\delta,s)}_\mathcal{X}(d) )$ when we have 
$W^{(1)} ( \mathcal{D}^{(\delta,s)}_\mathcal{X}(d) )> W^{(1)} ( 
\mathcal{C}_\mathcal{X}(d) )$.

\section{Further results on the minimum distance in a special 
case}\label{monomial}
In this section we assume that 
$K_1, \ldots, K_n$ are fields such 
that $K_1 \subset K_2 \subset \cdots \subset K_n \subset \F_q$.

We write $\aff(\F_q^n)$ for the affine group of $\F_q^n$, i.e.\ the 
transformations 
of
$\F_q^n$ of the type  $\balpha \longmapsto A \balpha + \boldsymbol{\beta}$, where 
$A \in GL(n,\F_q)$
and $\boldsymbol{\beta} \in \F_q^n$. 
\begin{definition}
The affine group associated to 
$\mathcal{X}$ is
\[
\aff(\mathcal{X}) =
\{ \varphi: \X \rightarrow \X \mid \varphi = \psi_{|_{\X}} \textrm{ with } 
\psi \in \aff(\F_q^n) \text{ and }
\psi(\mathcal{X})=\mathcal{X} \}.
\]
\end{definition}

Let
$\{e_1,\ldots ,e_n\} \subset \F_q^n$ be the canonical basis of $\F_q^n$,
since $e_1,\ldots ,e_n\in \X$ we get that for each $\varphi \in 
\aff(\mathcal{X})$ there exists only one $\psi \in \aff(\F_q^n)$ such 
that $\varphi = \psi_{|_{\X}}$.

\begin{lemma}\label{matrix}
Let $\psi \in \aff(\F_q^n)$ be given by 
$\balpha \longmapsto A \balpha + \bbeta$, where 
$$
A=
\left(
\begin{matrix}
a_{11} & \cdots & a_{1n} \\
\colon& \ddots & \colon \\
a_{n1} & \cdots & a_{nn}
\end{matrix}
\right)
\quad
\mbox{and}
\quad
\bbeta = 
\left(
\begin{matrix}
b_1 \\
\colon \\
b_n
\end{matrix}
\right),
$$
and let $\varphi = \psi_{|_{\X}}$.
Then 
$\varphi \in \aff(\X)$ if and only if the following conditions are satisfied:
\begin{enumerate}
\item[(i)]
for all $i , j \in \{1 , \ldots , n\}$,
$a_{ij} \in K_i$,  $b_j \in K_j$ and
if $K_i \subsetneqq K_j$ then
$a_{ij}=0$;
\item[(ii)] for all $i \leq j \in \{1 , \ldots , n\}$ such that
$K_{i-1} \subsetneqq K_i = K_j \subsetneqq K_{j+1}$
the square submatrix formed by entries $a_{uw}$ with
$i \leq u,w \leq j$ is invertible.
\end{enumerate}
\end{lemma}
\begin{proof}
Let $\psi : 
\balpha \longmapsto A \balpha + \bbeta \in \aff(\F_q^n)$
and suppose that 
$\psi_{|_{\X}} = \varphi \in \aff(\X)$. For 
$\balpha = 0$ we get $\varphi(0)=\bbeta \in \X$, which
implies $b_j \in K_j$ for all $j \in \{ 1,\ldots , n\}$.
We also get  that the transformation 
$\psi_0 : 
\balpha \longmapsto A \balpha \in \aff(\F_q^n)$ is such that 
$\varphi_0 = {\psi_0}_{|_{\X}} \in \aff(\X)$.

Let
$\{e_1,\ldots ,e_n\} \subset \F_q^n$ be the canonical basis of $\F_q^n$.
For any $j \in \{1,\ldots , n\}$ we get
$$
\psi_0(e_j) =
\left(
\begin{matrix}
a_{1j} \\ a_{2j} \\ \colon \\ a_{nj}
\end{matrix}
\right)
\in \X
$$
and so $a_{ij} \in K_i$ for all $i \in \{1,\ldots  , n\}$.

Let $i,j \in \{1,\ldots  , n\}$ such that $K_i \subsetneqq K_j$ (so in 
particular 
$j > i$) and choose
$\gamma_j \in K_j \backslash K_i$. From $\gamma_j e_j \in \X$ we get
$$
\psi_0(\gamma_j e_j) =
\left(
\begin{matrix}
\gamma_j a_{1j} \\ \gamma_j a_{2j} \\ \colon \\ \gamma_j a_{nj}
\end{matrix}
\right)
\in \X
$$
and, in particular, $\gamma_j a_{ij} \in K_i$  which  is only possible if 
$a_{ij}=0$.

Assume that $K_1 \subsetneqq K_n$ and 
let $i_0, \ldots, i_t$ be integers such that 
$0 = i_0 < i_1 < \cdots < i_t = n$, 
with $K_{i_{u} + 1} = \cdots = 
K_{i_{u + 1}}$ for all $u = 0, \ldots, t - 1$ and $K_{i_{u}} \subsetneqq 
K_{i_{u+1}}$ for all $u \in \{1, 
\ldots, t - 1\}$. Then
the matrix $A$ can be written as
$$
A=
\left(
\begin{matrix}
B_1     & 0         & 0    & \cdots & 0 \\
*         & B_2      & 0     & \cdots & 0 \\
*         &  *         & B_3 & \cdots & 0 \\
\colon & \colon & \colon & \colon & \colon \\
*         &  *         &   *       &  *         & B_t
\end{matrix}
\right)
$$
where for all $j = 1, \ldots, t$ the matrix $B_j$ is 
of size $(i_j - i_{j - 1})\times (i_j - i_{j - 1})$. 
Since 
$\det A = \det B_1 \cdot \det B_2 \cdots \det B_t$ and $\det A \neq 0$ we get
for all $j = 1, \ldots, t$
that $B_j$ is invertible with coefficients in $K_{i_j}$.
Conversely, if $(i)$ and $(ii)$ are satisfied then 
it is easy to see that $\varphi \in \aff(\X)$.
\end{proof}

The affine group $\aff(\X)$ acts over the set of polynomials
$\F_q[X_1,\ldots , X_n]$
in the following way. Let $f \in \F_q[X_1,\ldots , X_n]$,
$\varphi \in \aff(\X)$ and $\psi \in \aff(\F_q^n)$ such that
$\psi_{|_{\X}} = \varphi$. We define $f \circ \varphi \in \F_q[X_1,\ldots , 
X_n]$ as
$f \circ \varphi (X_1,\ldots , X_n) = f \left( \psi(X_1,\ldots ,X_n) \right)$, 
where
$(X_1,\ldots ,X_n)$ is written as a column vector.

\begin{definition}
We say that $f, \, g \in \F_q[X_1,\ldots,X_n]$ are $\mathcal{X}$-equivalent if 
there exists
$\varphi \in \aff(\mathcal{X})$
such that  $f =g \circ \varphi$.
\end{definition}

In \cite{car-neu-emmas-book} affine cartesian codes were studied as images of 
polynomial 
functions 
evaluated in the points of $\X$, and two polynomials define the same function 
if their difference belongs to $I_\X$. In the following result we rewrite
\cite[Thm.\ 3.5]{car-neu-emmas-book} without using the function concept.

\begin{theorem}
\label{pesominimofinal}
Let 
$d = \displaystyle \sum_{i=1}^{k} (d_i - 1)+ \ell $, $0\le  k < n$ and
$0 < \ell \leq d_{k+1} - 1$, 
the minimal weight codewords of 
$C_\mathcal{X}(d)$ are of the form $\Psi(f)$ where
$f \in \F_q[X_1,\ldots , X_n]_{\leq d}$
is such that there exists $g \in \F_q[X_1,\ldots , X_n]$, with $f - g \in I_\X$ 
and $g$ is $\mathcal{X}$-equivalent to a polynomial 
$$
h = \sigma \prod_{i=1,i \neq j}^{k+1} (X_i^{d_i-1} - 1)
\prod_{t=1}^{d_j - (d_{k+1}-\ell)} 
\left(
X_j - \alpha_t
\right)
\, ,
$$
where $j \in \{1, \ldots, k + 1\}$  is such that $d_j - (d_{k+1}-\ell) \geq 0$,
$\sigma \in \F_q^*$ and $\alpha_1, \ldots, \alpha_{d_j - (d_{k+1}-\ell)}$
are distinct elements of $K_j$ (if $d_j - (d_{k+1}- \ell) = 0$ we take the 
second 
product as being equal to 1).
\end{theorem}

The following result describes a property of certain polynomials of degree 1 
which will be used in the next proposition.

\begin{lemma}\label{linear}
Let $p = \gamma_1 X_1 + \cdots +  \gamma_h X_h + \eta \in \F_q[X_1,\ldots , 
X_n]$, where $\gamma_1, \ldots, \gamma_h \in  \F_q$ and 
$\gamma_h \neq 0$. Then
there exists $\varphi \in \aff(\X)$ and $j \in \{1,\ldots , n\}$
such that $X_j \circ \varphi=p$ if and only if
$\gamma_i \in K_j$ for all $i \in \{1,\ldots , h\}$, $\eta \in K_j$ and $K_h = 
K_j$.
\end{lemma}
\begin{proof}
Assume that there exists $\varphi \in \aff(\X)$ such that $X_j \circ \varphi = 
p$ 
for some 
$j \in \{1,\ldots , n\}$, and let $\psi \in \aff(\F_q^n)$ be such that 
$\varphi = \psi_{|_{\X}}$. If $\psi$ is given by 
$\balpha \longmapsto A \balpha + \bbeta$,
then the $j$-th line of $A$
has to be  $(\gamma_1, \ldots, \gamma_h, 0, \ldots, 0)$, so $\gamma_i \in K_j$ 
for all $i 
\in \{1, \ldots, h\}$, likewise the $j$-th entry of $\bbeta$ has to be $\eta$, 
so 
that $\eta \in K_j$. From the general form of $A$, which was described in Lemma 
\ref{matrix} we get that  $K_h 
= K_j$. The proof of the converse is simple and follows from Lemma \ref{matrix}.
\end{proof}

\begin{definition}
A linear form $L = \gamma_1 X_1 + \cdots +  \gamma_h X_h$, where $\gamma_h \neq 
0$, $\gamma_i \in K_j$ for all $i \in \{1,\ldots , h\}$ and 
$K_h = 
K_j$ will be called
a $\X$-linear form over $K_j$.
\end{definition}

\begin{proposition}\label{lider}
Let $f \in \F_q[X_1,\ldots X_n]$ be a polynomial of degree 
$$
d = \sum_{i=1}^k (d_i-1) + \ell,
$$
where  $ 0 \leq k <n$ and $0 < \ell \leq d_{k+1}-1$.
Assume that no monomial in $f$ is a multiple of $X_i^{d_i}$, for all $i = 1, \ldots, n$.
If
$w(\Psi(f)) = \displaystyle (d_{k+1}-\ell)\prod_{i=k+2}^n d_i$ then there exists 
a monomial in $f$ of the form 
$\displaystyle X_{t_{j}}^{d_j- (d_{k+1} - \ell)}
\prod_{\substack{i=1\\ i \neq j}}^{k+1} X_{t_i}^{d_i - 1}$
for some $1 \le j \le k+1$ such that $d_j \geq d_{k+1}- \ell$,
where $t_1, \ldots, t_{k+1}$ are distinct elements of  $\{1, \ldots, n\}$
and 
$K_{t_i}=K_i$ for all $ i \in \{1,\ldots , k+1\}$.
	%	a polynomial $g \in 
	%	\F_q[X_1,\ldots X_n]$, of degree $e$, such that $\psi_e(g) = \psi_e(f)$ 
	%	and,  after a permutation of the variables, if necessary, the coefficient 
	%	of the monomial 
	%	$X_1^{q-1} \cdots X_{k}^{q-1} X_{k+1}^{\ell}$ in $g$ is not zero.
\end{proposition}
\begin{proof}
From \eqref{md-cartesian} we get that $w(\Psi(f)) = W^{(1)} 
(\mathcal{C}_\mathcal{X}(d))$ so from
Theorem \ref{pesominimofinal} there exist 
$j \in \{ 1,\ldots , k+1\}$ and
a polynomial
$$ 
g = \sigma
\prod_{\substack{i=1\\ i \neq j}}^{k+1} ( (L_i - \alpha_i)^{d_i-1} - 1 )
\prod_{s=1}^{d_j- (d_{k+1} - \ell)} (L_{j} - \beta_s ),
$$
where $\sigma \in \F_q$, $\beta_1, \ldots, \beta_{d_j- (d_{k+1} - \ell)}$ are 
distinct elements of 
$K_j$,
$L_i$ is a $\X$-linear form over $K_i$ and $\alpha_i \in K_i$ for all $i \in 
\{1, \ldots, k+1\}$, the forms  
$L_1, \ldots, L_{k + 1}$ are linearly independent over $\F_q$, and
$f - g \in I_{\X}$.
Since $\deg(g) = d$
we get $g \in 
\F_q[X_1,\ldots X_n]_{\leq d}$ and
$\Psi(g) = \Psi(f) \in \mathcal{C}_\mathcal{X}(d)$.

Assume, for a moment, that there are at least two factors in the first product 
in the definition of $g$, i.e. assume that there exist $u,w \in \{1, \ldots, 
k+1\}$ with $u<w$ 
and $u,w \neq 
j$.
Observe that
evaluating 
the polynomial 
$$
( (L_u - \alpha_u)^{d_u-1} - 1 )( (L_w - \alpha_w)^{d_w-1} - 1 )
$$
at the points of  $\X$  we get the value zero, except for those $P \in \X$
where $L_u(P) = \alpha_u$ and $L_w(P) = \alpha_w$, and at these points we get 
$1$.
For any
$\gamma \in K_w$
we get the same results evaluating the polynomial 
$$
( (L_u - \alpha_u)^{d_u-1} - 1 )
( (L_w - \alpha_w -\gamma (L_u - \alpha_u))^{d_w-1} - 1 )
$$
at the points of $\X$.
Thus we may replace, in the polynomial $g$, the factor 
$(L_w - \alpha_w)^{d_w-1} - 1$ by the factor $(L_w - \alpha_w -\gamma (L_u - 
\alpha_u))^{d_w-1} - 1$ obtaining 
a polynomial $\tilde{g}$ such that $\Psi(\tilde{g}) = \Psi(g)$, and a fortiori  
$\tilde{g} - g \in I_{\X}$. This reasoning shows that we may perform a Gaussian 
elimination process in the set $\{L_i - \alpha_i \mid i = 1, \ldots, k+1, i 
\neq j\}$, 
starting with the linear form with the greatest index and proceeding to the 
linear form with the least index, 
and find a set of $k$ integers  $1 \leq t_1 < \cdots < t_{j-1}<t_{j+1}<\cdots < 
t_{k+1} \leq n$ such that after the elimination process 
we may assume that 
$L_i = X_{t_i} +\sum_{w < t_i, w \notin A} a_{iw} X_w$ for all
$i \in \{1,\ldots , k+1 \} \backslash \{j\}$, where $A = \{t_i \mid i = 1, 
\ldots, j - 1, j+1, \ldots, k+1\}$. Observe that 
$K_{t_i}=K_i$ for all $ i \in \{1,\ldots , k+1\}\setminus\{j\}$ and 
we still have  
\[ 
f - \tau
\prod_{\substack{i=1\\ i \neq j}}^{k+1} ( (L_i - \gamma_i)^{d_i-1} - 1 )
\prod_{s=1}^{d_j- (d_{k+1} - \ell)} (L_{j} - \beta_s ) \in 
I_{\X}
\]
for some $\tau \in \F_q$, and $\gamma_i \in K_i$ for all $i \in \{1, \ldots, 
k+1\} \setminus \{j\}$.

Let $i \in \{1,\ldots , k+1 \} \backslash \{j\}$ be such that $K_{t_i} \subset 
K_j$ and let $\xi \in K_j$, then the polynomials 
\[
( (L_i - \gamma_i)^{d_i-1} - 1 ) \prod_{s=1}^{d_j-(d_{k+1} - \ell)} (L_{j} - 
\beta_s )
\]
	and 
\[
( (L_i - \gamma_i)^{d_i-1} - 1 )
\prod_{s=1}^{d_j-(d_{k+1} - \ell)} (L_{j} - \beta_s - \xi (L_i - \gamma_i)  )
\]	
yield the same value when evaluated at any $P \in \X$, so their difference is 
in $I_{\X}$.
As before, after a Gauss-Jordan elimination process, we may assume that 
$L_{j}=X_{t_{j}} +\sum_{w < t_{j}, w \notin A} a_{j,w} X_w$, with
$t_j \notin A$ and $K_{t_j} = K_j$. Again, 
\[ 
f - \eta
\prod_{\substack{i=1\\ i \neq j}}^{k+1} ( (L_i - \gamma_i)^{d_i-1} - 1 )
\prod_{s=1}^{d_j- (d_{k+1} - \ell)} (L_{j} - \theta_s ) \in 
I_{\X}.
\]
still holds, for some $\eta \in \F_q$ and $\theta_s \in K_j$, $s = 1, \ldots, 
d_j - (d_{k+1} - \ell)$.
Taking the 
lexicographic order where $X_1 < \cdots < X_n$, we get that 
the leading monomial of the right hand side polynomial in the above difference 
is 
$$M =  X_{t_{j}}^{(d_j - (d_{k+1} - \ell)}
\prod_{\substack{i=1\\ i \neq j}}^{k+1} X_{t_i}^{d_i - 1}.
$$
Since the remainder 
in the division of $f - g$ by the Gr\"obner basis
$\{X_1^{d_1} - X_1,\ldots, X_n^{d_n} - X_n\}$ is 
zero, and $M$ is not a multiple of 
$X_i^{d_i}$ for all 
$i = 1,\ldots, n$, we get that this monomial must also appear in $f$. 
\end{proof}

We apply the above result to obtain a converse to Theorem \ref{min_dist_1}.

\begin{theorem}\label{secondweight}
Assume that 
$K_1, \ldots, K_n$ are fields such 
that $K_1 \subset K_2 \subset \cdots \subset K_n \subset \F_q$.
Let $d =\displaystyle \sum_{i=1}^k (d_i-1) + \ell$ where  $ 0 \leq k <n$ and $0 < \ell \leq d_{k+1}-1$.
If 
$W^{(1)} ( \mathcal{D}^{(\delta,s)}_\mathcal{X}(d) )
= W^{(1)} ( \mathcal{C}_\mathcal{X}(d) )$
then one of the following conditions must hold:
\begin{enumerate}
	\item[(i)] $k+2\leq n$ and $d_{k+2} \leq d_s$;
	\item[(ii)] $d_s \leq d_{k+1}$ and $0 \leq  d_s - (d_{k+1} - \ell) < r$.
\end{enumerate}
\end{theorem}
\begin{proof}
Suppose that condition 
$(i)$ is not satisfied, then 
$n=k+1$ or $d_s < d_{k+2}$,
which implies, in both cases, that $d_s \leq d_{k+1}$.
If condition (ii) is also not satisfied we must then have 
$d_s - (d_{k+1} - \ell) < 0$ or $r \leq d_s - (d_{k+1} - \ell)$.
Thus if conditions (i) and (ii) 
are not satisfied, then $n=k+1$ or $d_s < d_{k+2}$, and 
$d_s - (d_{k+1} - \ell) < 0$ or $d_s - (d_{k+1} - \ell) \geq r$.

We assume that $W^{(1)} ( \mathcal{D}^{(\delta,s)}_\mathcal{X}(d) )
= W^{(1)} ( \mathcal{C}_\mathcal{X}(d) )$ holds, and 
let $f \in \Pp^{(\delta, s)}_d$ be a polynomial of degree
$d$ such that
$$
w(\Psi(f))=W^{(1)} ( \mathcal{C}_\mathcal{X}(d)).
$$
From
Proposition \ref{lider}
there exists 
a monomial in $f$ of the form 
$$
M_j=
\displaystyle X_{t_{j}}^{d_j- (d_{k+1} - \ell)}
\prod_{\substack{i=1\\ i \neq j}}^{k+1} X_{t_i}^{d_i - 1}
$$
for some $1 \le j \le k+1$ such that $d_j \geq d_{k+1}- \ell$,
where $t_1, \ldots, t_{k+1}$ are distinct elements of  $\{1, \ldots, n\}$
and 
$K_{t_i}=K_i$ for all $ i \in \{1,\ldots , k+1\}$.

If $n=k+1$ then $\{t_1, \ldots, t_{k+1}\}=\{1,\ldots , k+1\}$, which implies that
$s=t_i$ for some $ i \in \{1,\ldots , k+1\}$. 
If $\deg_{X_{s}} M_j = d_s - 1$ then
$\deg_{X_{s}} M_j \geq d_s - \delta + 1$, and if 
$\deg_{X_{s}} M_j = d_s- (d_{k+1} - \ell)$ then we cannot have
$d_s - (d_{k+1} - \ell) < 0$ so we must have $d_s - (d_{k+1} - \ell) \geq r$,
which leads to a contradiction since we also must have 
$\deg_{X_{s}} M_j < d_s - \delta + 1 = r$.

Thus we suppose now that $k+1<n$ and $d_s<d_{k+2}$. Let $u$ be the integer such 
that
$s < u \leq k+2$ and $d_{u-1}<d_u = d_{k+2}$.
From the definition of the set $\{t_1,\ldots ,t_{k+1}\}$ we have, in 
particular, that 
$K_{t_i}=K_i$ for all $ i \in \{1,\ldots , u-1\}$, so
$s = t_i$ for some $ i \in \{1,\ldots , u-1\} \subset \{1,\ldots , k+1\}$. As 
above, analysing the degree of $M_j$ we get  
$f \notin \Pp^{(\delta, s)}_d$, which finishes the proof.
\end{proof}

Thus, if conditions (i) and (ii) of the above theorem are not satisfied, 
then from Theorem 
\ref{secondweight}
we get that
$W^{(1)}(\mathcal{D}^{(\delta,s)}_\mathcal{X}(d)) > 
W^{(1)}(\mathcal{C}_\mathcal{X}(d))$. Since  
$\mathcal{D}^{(\delta,s)}_\mathcal{X}(d)\subset\mathcal{C}_\mathcal{X}(d)$
we must have 
$W^{(1)}(\mathcal{D}^{(\delta,s)}_\mathcal{X}(d)) \geq 
W^{(2)}(\mathcal{C}_\mathcal{X}(d))$ where 
$W^{(2)}(\mathcal{C}_\mathcal{X}(d))$ denotes the second lowest codeword weight 
in $\mathcal{C}_\mathcal{X}(d)$, also called next-to-minimal weight of 
$\mathcal{C}_\mathcal{X}(d)$. 
The values for $W^{(2)} ( \mathcal{C}_\mathcal{X}(d))$ were determined in the 
series of papers \cite{carvalho}, \cite{car-neu2017} and \cite{car-neu2020b}. 
These papers contain, in particular, the values for the special case where 
$\mathcal{X} = \F_q^n$, which had already been determined by a combination of 
results by 
several authors -- the reader may find a historical survey of these results in 
\cite{car-neu2020a}. From these papers, we get that, writing  
$d =\displaystyle \sum_{i=1}^k (d_i-1) + \ell$ where  $ 0 \leq k <n$ and $0 < 
\ell \leq d_{k+1}-1$, 
the values for 
$W^{(2)} ( \mathcal{C}_\mathcal{X}(d))$ are as follows:
\begin{enumerate}
\item if $n=k+1$ then
(see \cite[Theorem 2.6]{carvalho})
$$
W^{(2)} ( \mathcal{C}_\mathcal{X}(d) )= d_n - \ell +1;
$$
\item if $3 \leq d_1 \leq \cdots \leq d_n$ and either $\ell = 1$ and  
$d_{k+1}<d_{k+2}$, or $\ell \geq 2$ 
then
(see \cite[Theorem 3.10]{car-neu2017})
$$
W^{(2)} ( \mathcal{C}_\mathcal{X}(d))=
(d_{k+1}-\ell+1 )(d_{k+2}-1) \prod_{i=k+3}^n d_i ;
$$
\item if $4 \leq d_{i}=q$ for all $i \in \{1,\ldots , n \}$ and $\ell=1$
then 
(see e.g.\ \cite[Theorem 3.5]{car-neu2020b})
$$
W^{(2)} ( \mathcal{C}_\mathcal{X}(d))=q^{n-k};
$$
\item For all other cases where $d_{k+1}=d_{k+2}$, $\ell=1$
and $3 \leq d_1 \leq \cdots \leq d_n$
then
(see \cite[Theorem 3.5]{car-neu2020b})
$$
W^{(2)} ( \mathcal{C}_\mathcal{X}(d))=
(d_{k+1}^2-1) \prod_{i=k+3}^n d_i.
$$
\end{enumerate}

\begin{corollary}\label{last-result}
Assume that $n=k+1$ or
$3 \leq d_1 \leq \cdots \leq d_n$, if the conditions
$(i)$ and $(ii)$ of the above proposition are not satisfied
and
$d_s - (d_{k+1} - \ell)=r$ then
$$
W^{(1)} ( \mathcal{D}^{(\delta,s)}_\mathcal{X}(d) )
=
\left\{
\begin{matrix}
d_n - \ell +1 & \text{if } n=k+1 ; \\
(d_{k+1}-\ell+1 )(d_{k+2}-1) \prod_{i=k+3}^n d_i
& \text{if } n>k+1 .
\end{matrix}
\right.
$$
\end{corollary}

\begin{proof}
If (i) and (ii) of Theorem \ref{secondweight} are 
not satisfied, then, as in the above proof we get that 
 $n=k+1$ or $d_s < d_{k+2}$, and 
$d_s - (d_{k+1} - \ell) < 0$ or $d_s - (d_{k+1} - \ell) \geq r$.
These last two inequalities we replace by the hypothesis $d_s - (d_{k+1} - 
\ell)=r$. 

Let 
$$
g = \prod_{\substack{i=1\\ i\neq s}}^{k+1} (X_i^{d_i-1}-1)
\cdot \prod_{h=1}^{d_s - (d_{k + 1} - \ell) - 1} (X_{s}-\beta_h),
$$
then $\deg(g) = \sum_{i=1, \, i\neq s}^{k+1} (d_i - 1)  + 
d_s - (d_{k + 1} - \ell) - 1 = \sum_{i = 1}^k(d_i - 1) + \ell - 1 = d - 1$
(if $d_s - (d_{k+1} - \ell) = 1$ then we take  the second product in the
definition of $g$ as being 1 and still get $\deg(g) = d - 1$). 
Clearly  $g \in \Pp^{(\delta, s)}_d$ since $\deg_{X_{s}} g < r$.

Suppose that $n = k+1$,
 since 
$w(\Psi(g))= d_{k+1} - \ell +1$ we must have 
$W^{(1)} ( \mathcal{D}^{(\delta,s)}_\mathcal{X}(d) )= 
W^{(2)}(\mathcal{C}_\mathcal{X}(d))$ (from the above data on 
$W^{(2)}(\mathcal{C}_\mathcal{X}(d))$.

We now treat the case where $k+1 < n$, then we have $d_s < d_{k+2}$, and 
from the hypothesis we also have $3 \leq d_1 \leq \cdots \leq d_n$. Assume that 
$d_{k+1} < 
d_{k+2}$ and let $f = g . X_{k+2}$, then 
$\deg(f) = d$ and $f \in \Pp^{(\delta, s)}_d$, from 
$w(\Psi(f))= (d_{k+1}-\ell+1 )(d_{k+2}-1) \prod_{i=k+3}^n d_i$ we get  
$W^{(1)} ( \mathcal{D}^{(\delta,s)}_\mathcal{X}(d) )= W^{(2)}
(\mathcal{C}_\mathcal{X}(d) )$. In the case where $d_{k+1} = d_{k+2}$ from  
$d_s < d_{k+2}$ and $d_s - (d_{k + 1} - \ell) = \ell - (d_{k+2} - d_s) = r \geq 
1$ we see that we must have $\ell \geq 2$,  so again we have 
$w(\Psi(f))=  W^{(2)} ( 
\mathcal{C}_\mathcal{X}(d) )$, which finishes the proof.
\end{proof}

\section{Examples}
In this section we present some tables with numerical data obtained from the 
above results. In the tables, we use the following notation: $m=|\X|$ is the 
length of $\mathcal{D}^{(\delta,s)}_{\mathcal{X}}(d)$, $\kappa = \dim_{\F_q} 
\mathcal{D}^{(\delta,s)}_{\mathcal{X}}(d)$, 
$v = W^{(1)}(\mathcal{C}_\mathcal{X}(d))$,
$w =W^{(1)} ( \mathcal{D}^{(\delta,s)}_{\mathcal{X}}(d) )$ and we denote by 
$N=m - \kappa -
\left( \left\lceil   \dfrac{\kappa}{r} \right\rceil - 1 \right)
(\delta -1) + 1$ 
the upper bound for the minimum distance, which appears in 
Theorem \ref{min_dist_1}. In the tables $d$ runs in the range $1 \leq d \leq 
\tilde{d}$.  When $w \neq v$ then $w \geq W^{(2)}(\mathcal{C}_\mathcal{X}(d))$
and in Section 5 the values for $W^{(2)}(\mathcal{C}_\mathcal{X}(d))$
are presented. Yet, when $w \neq v$ and we are in the hypotheses of  	
Corollary \ref{last-result}, then we write the true value of $w$.

In the table below, for the $d$ presented we always have $w = v$.

\begin{table}[htbp]
	\centering
	\begin{tabular}{c|ccccccccccc}
		\multicolumn{12}{c}{$\mathcal{D}^{(25,2)}_{\mathcal{X}}(d)$} \\
		\hline
		$d$  & $4$ & $5$   &$10$&$15$&$20$&$25$&$26$ & $27$ & $28$ & $29$ & 
		$30$  \\
		\hline
		$m$  & $343$& $343$& $343$&$343$& $343$ & 
		$343$&$343$&$343$&$343$&$343$&$343$  \\
		\hline
		$\kappa$&$15$&$21$&$56$&$91$&$126$&$160$&$165$&$169$&$172$&$174$&$175$  
		\\
		\hline
		$w$    & $147$& $98$&$45$   & $40$  &$35$  & $30$&$29$& 
		$28$&$27$&$26$&$25$ \\
		\hline
		$N$     & $329$ & $323$ & $240$&$181$& 
		$98$&$40$&$35$&$31$&$28$&$26$&$25$
	\end{tabular}
	\caption{$\X := \F_7 \times \F_{49}$}
	\label{caseq749}
\end{table}

In the table below, for some values of $d$ we have $w \neq v$.

\begin{table}[htbp]
	\centering
	\begin{tabular}{c|ccccccccc}
		\multicolumn{10}{c}{$\mathcal{D}^{(4,1)}_{\mathcal{X}}(d)$} \\
		\hline
		$d$  &  $2$   &$3$&$24$&$25$&$26$&$27$ & $47$ & $48$ & $49$  \\
		\hline
		$m$  & $3125$& $3125$&$3125$& $3125$ & 
		$3125$&$3125$&$3125$&$3125$&$3125$  \\
		\hline
		$\kappa$&$9$&$16$& $625$&$674$&$721$&$766$&    $1246$&$1249$&$1250$  \\
		\hline
        $v$    & $1875$&$1250$&   $125$&$100$
         &$75$&$50$& 
         $6$&$5$&$4$ \\
		\hline
		$w$    & $2400$&$\geq 1800$& $125$&$100$
		&$96$&$\geq 72$& 
		$\geq 7$&$5$&$4$ \\
		\hline
		$N$ &$3105$&$3089$&  $1565$&$1444$&$1325$&$1214$&
		$14$&$5$&$4$
	\end{tabular}
	\caption{$\X :=  \F_5 \times \F_{25} \times \F_{25}$}
	\label{caseq52525}
\end{table}

\bibliographystyle{plain}

\end{document}